%% file: prl_bsm_v5_0.tex
\documentclass[aps,prl,twocolumn,showpacs,groupedaddress]{revtex4}  
\usepackage{graphicx}  
\usepackage{dcolumn}   
\usepackage{bm}        
\usepackage{amssymb}   

\newcommand{\dzer}         {D\O}

\newcommand{\ds}        {\ensuremath{{D_{s}^{-}}}}

\newcommand{\bs}        {\ensuremath{{B_{s}^{0}}}}

\begin{document}


\hspace{5.2in} \mbox{Fermilab-Pub-06/055-E}

\title{Direct Limits on the $B^0_s$ Oscillation Frequency}
\input list_of_authors_r2.tex  
\date{submitted to PRL on 15 March 2006; published 14 July 2006}

\begin{abstract}
We report results of a study of the $B^0_s$ oscillation frequency
using a large sample of $B^0_s$ semileptonic decays corresponding
to approximately 1~fb$^{-1}$ of integrated luminosity collected
by the \dzer\ experiment at the Fermilab
Tevatron Collider  in 2002--2006.
The amplitude method gives a lower limit on the $B^0_s$ oscillation
frequency at 14.8~ps$^{-1}$ at the 95\% C.L. 
At $\Delta m_s = 19$~ps$^{-1}$, the amplitude deviates from
the hypothesis $\mathcal{A}=0$ ($\mathcal{A}=1$) by 2.5 (1.6) 
standard deviations, corresponding to a two-sided C.L. of 1\% (10\%).
A likelihood scan over the oscillation frequency, $\Delta m_s$, gives
a most probable value of 19~ps$^{-1}$ and a range of $17 < \Delta m_s < 21$~ps$^{-1}$ 
at the 90\% C.L., assuming Gaussian uncertainties.
This is the first direct two-sided bound measured by a single experiment.
If $\Delta m_s$ lies above 22~ps$^{-1}$, then the probability that it
would produce
a likelihood minimum similar to the one observed in the interval
$16 < \Delta m_s < 22$~ps$^{-1}$ is $(5.0\pm0.3)$\%. 

\end{abstract}

\maketitle 

Measurements of flavor oscillations in the $B^0_d$ and
$B^0_s$ systems provide important 
constraints on the CKM unitarity triangle and the
source of CP violation in the standard model (SM)~\cite{pdg}. 
The phenomenon of $B^0_d$ oscillations 
is well established~\cite{hfag}, with
a precisely measured oscillation frequency $\Delta m_d$. In 
the SM, this parameter is proportional to
the combination
$|V^*_{tb} V_{td}|^2$ of CKM matrix elements. 
Since the matrix element $V_{ts}$ is larger than $V_{td}$,
the expected frequency $\Delta m_s$ 
is higher. As a result, $B^0_s$ oscillations
have not been observed by any previous experiment and 
the current 95\% C.L. lower limit on
$\Delta m_s$ is 16.6~ps$^{-1}$~\cite{hfag}.
A measurement of 
$\Delta m_s$ would yield the ratio
$|V_{ts}/V_{td}|$, which has a smaller uncertainty than $|V_{td}|$ 
alone due to 
the cancellation of certain theory uncertainties.
If the SM is correct, and if current limits 
on $B^0_s$ oscillations are not included, then global fits
to the unitarity triangle favor 
$\Delta m_s = 20.9 ^{+4.5}_{-4.2}$~ps$^{-1}$~\cite{ckm_fit} or $\Delta m_s = 21.2 \pm3.2$~ps$^{-1}$~\cite{ut_fit}.

In this Letter, we present
a study of $B^0_s$-$\bar{B}^0_s$ oscillations
carried out using semileptonic $B^0_s \rightarrow \mu^+ D^-_s X$ decays~\cite{charge_conj}
collected by the \dzer\ experiment at Fermilab in $p\bar{p}$
collisions at $\sqrt{s} = 1.96$~TeV. 
In the $B^0_s$-$\bar{B}^0_s$ system there are two
mass eigenstates, the heavier (lighter) one having mass
$M_H$ ($M_L$) and decay width $\Gamma_H$ ($\Gamma_L$). 
Denoting $\Delta m_s = M_H - M_L$,
$\Delta \Gamma_s = \Gamma_L - \Gamma_H$, 
$\Gamma_s = (\Gamma_L + \Gamma_H)/2$, 
the time-dependent
probability $P$ that an initial $B^0_s$ decays at time $t$ 
as $B^0_s \rightarrow \mu^+ X$ ($P^{\mathrm{nos}}$)
or $\bar{B}^0_s \rightarrow \mu^- X$ ($P^{\mathrm{osc}}$) is given by 
$P^{\mathrm{nos}/\mathrm{osc}} = e^{-\Gamma_s t}(1 \pm \cos\Delta m_s t)/2$,
assuming that $\Delta\Gamma_s/\Gamma_s$ is small and neglecting CP violation.
Flavor tagging a $b$ ($\bar{b}$) on the
opposite side to the signal meson establishes the 
signal meson as a $B^0_s$ 
($\bar{B}^0_s$) at time $t=0$. 

The \dzer\ detector is described in detail elsewhere~\cite{run2det}.
Charged particles are reconstructed using the central tracking system which 
consists
of a silicon microstrip tracker (SMT) and a central fiber tracker
(CFT), both located within a 2-T superconducting solenoidal magnet.
Electrons are identified by the preshower and liquid-argon/uranium 
calorimeter. Muons are identified by the muon system which
 consists of a layer of
tracking detectors and scintillation trigger counters in front of
1.8-T iron toroids, followed by two similar layers after the toroids~\cite{run2muon}.

No explicit trigger requirement was made, although most of
the sample was collected with single muon triggers.
The decay chain
$\bs\rightarrow
\mu^{+}D_{s}^{-}X$, $D_{s}^{-}\rightarrow \phi \pi^{-}$,
$\phi\rightarrow K^{+}K^{-}$ was then reconstructed.
The charged tracks were required to have signals in both the CFT and SMT.
Muons were required to have 
transverse momentum $p_T(\mu^+) > 2$~GeV$/c$ and
momentum $p(\mu^+) > 3$~GeV$/c$, 
and to have measurements in at least two layers of the muon system.
All  charged tracks in the event were clustered into
jets~\cite{durham}, and the $D^-_s$ candidate was reconstructed
from three tracks found in the same jet as the reconstructed muon.
Oppositely charged particles with 
$p_T > 0.7$~GeV$/c$ were assigned
the kaon mass  and were required to have an invariant
mass $1.004 < M(K^+ K^-) < 1.034$~GeV$/c^2$, consistent with that of a
$\phi$ meson. The third track 
was required to have 
$p_T > 0.5$~GeV$/c$ and
charge 
opposite to that of 
the muon charge and was assigned the pion mass.  
The three tracks were required to form a common $D_s^-$ vertex
using the algorithm described in Ref.~\cite{vertex}.
To reduce combinatorial background, the $D^-_s$ vertex was
required to have a positive displacement in the transverse plane, 
relative to the $p\bar{p}$ collision point (or primary vertex, PV), 
with at least $4\sigma$ significance.
The cosine of the angle between the $D^-_s$ momentum and the 
direction from the PV to the $D^-_s$ vertex
was required to be greater than 0.9.
The trajectories of the muon and $D^-_s$ candidates were
required to originate from a common $B^0_s$ vertex, and
the $\mu^+ D_s^-$ system was required to have an invariant mass
between 2.6 and 5.4~GeV$/c^2$.

\begin{figure}
\includegraphics[width=0.48\textwidth]{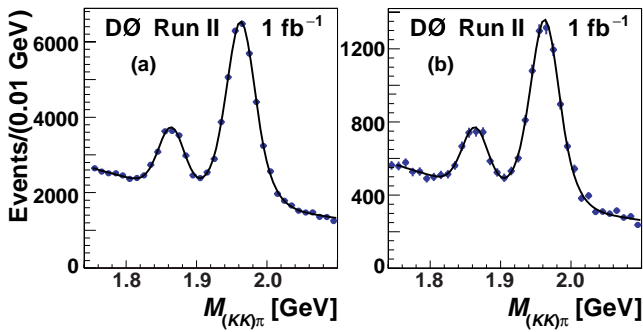}
\caption{\label{prl_fig1} 
$(K^+K^-)\pi^-$ invariant mass distribution for
(a) the untagged $B^0_s$ sample, and (b) for candidates
that have been flavor-tagged.  The left and right peaks correspond to
$\mu^+ D^{-}$ and $\mu^+ D^-_{s}$ candidates, respectively. The curve
is a result of fitting a signal plus background model to the data. 
}
\end{figure}

To further improve $B^0_s$ signal selection, 
a likelihood ratio method~\cite{like_ratio}
was utilized. Using $M(K^+K^-\pi)$ sideband ($B$) and
sideband-subtracted signal ($S$) distributions in the data, probability
density functions ({\sl pdf}s) were found for a number of 
discriminating variables:
the helicity angle between the
$D_s^-$ and $K^{\pm}$ momenta in the $\phi$ center-of-mass frame,
the isolation of the $\mu^+ D_s^-$ system,  
the $\chi^2$ of the 
$D_s^-$ vertex, the
invariant masses $M(\mu^+ D_s^-)$ 
and $M(K^+ K^-)$, and  $p_T(K^+ K^-)$.
The final requirement on the combined
selection likelihood ratio variable, $y_{\mathrm{sel}}$, was chosen to maximize
the predicted ratio $S/\sqrt{S+B}$.
The total number of $D_s^-$ candidates after these
requirements was 
$N_{\mathrm{tot}} = 26,\!710 \pm 556 \thinspace \mathrm{(stat)}$, as shown 
in Fig.~\ref{prl_fig1}(a).

The performance of the opposite-side
flavor tagger (OST)~\cite{OST} is
characterized by the efficiency
$\epsilon = N_{\mathrm{tag}}/N_{\mathrm{tot}}$, where 
$N_{\mathrm{tag}}$ is the number of tagged $B^0_s$ mesons;
tag purity $\eta_s$, defined as 
$\eta_s = N_{\mathrm{cor}}/N_{\mathrm{tag}}$, where
$N_{\mathrm{cor}}$ is the number of $B^0_s$ mesons with correct
flavor identification; and the dilution $\mathcal{D}$, related to purity as 
$\mathcal{D} \equiv 2\eta_s -1$.
Again, a likelihood ratio method was used.
In the construction of the flavor discriminating variables $x_1,
..., x_n$ for each event,
an object, either a lepton $\ell$ (electron or muon) or a reconstructed
secondary vertex (SV), was defined to be on the opposite side from
the $B^0_s$ meson if
it satisfied 
$\cos\varphi(\vec{p}_{\ell~{\mathrm{or~SV}}}, \vec{p}_B) < 0.8$,
where $\vec{p}_B$ is the reconstructed three-momentum of the 
$B^0_s$ meson, and $\varphi$ is the azimuthal angle about the beam axis.  
A lepton jet charge was formed
as $Q^{\ell}_J = \sum_i q^i p^i_T/\sum_i p^i_T$, where
all charged particles are summed, including the lepton,
inside a cone
of $\Delta R = \sqrt{(\Delta\varphi)^2 + (\Delta\eta)^2} < 0.5$ centered
on the lepton.
The SV charge
was defined as 
$Q_{\mathrm{SV}} = \sum_i (q^i p_L^i)^{0.6}/\sum_i (p^i_L)^{0.6}$,
where
all charged 
particles associated with the SV are summed,
and $p^i_L$ is the longitudinal momentum of track $i$
with respect to the
direction of the SV momentum. Finally,
event charge is defined as 
$Q_{\mathrm{EV}} =  \sum_i q^i p^i_T/\sum_i p^i_T$,
where the sum is over all tracks with $p_T > 0.5$~GeV$/c$
outside a cone of $\Delta R > 1.5$ centered on the $B^0_s$ direction.
The {\sl pdf} of each discriminating variable was found
for $b$ and $\bar{b}$ quarks
using a large data sample of $B^+ \rightarrow \mu^+ \nu \bar{D}^0$ 
events where the initial state is 
known from the charge of the decay muon.

For an initial $b$ ($\bar{b}$) quark, the {\sl pdf} for a given 
variable $x_i$ is denoted 
$f_i^b(x_i)$ ($f_i^{\bar{b}}(x_i)$),
and the combined tagging
variable is defined as
$d_{\mathrm{tag}} = (1-z)/(1+z)$, where $z=\prod_{i=1}^{n}(f_i^{\bar{b}}(x_i)/f_i^{b}(x_i))$.
The variable $d_{\mathrm{tag}}$ varies between $-1$ and $1$.
An event with $d_{\mathrm{tag}} > 0$~$(< 0)$ is tagged as a 
$b$ ($\bar{b}$) quark.


The OST purity was determined from 
large samples of $B^+ \rightarrow \mu^+ \bar{D}^0 X$ 
(non-oscillating) and 
$B^0_d \rightarrow \mu^+ D^{*-} X$ (slowly oscillating) semileptonic
candidates. 
An average value of
$\epsilon \mathcal{D}^2 =
[2.48 \pm 0.21 \thinspace \mathrm{(stat)} 
          ^{+0.08}_{-0.06} \thinspace \mathrm{(syst)}]\%$ was
obtained~\cite{OST}.
The estimated event-by-event dilution as a function
of $|d_{\mathrm{tag}}|$ was determined by 
measuring $\mathcal{D}$
in bins of 
$|d_{\mathrm{tag}}|$ and parametrizing 
with a third-order polynomial for
$|d_{\mathrm{tag}}| < 0.6$. For $|d_{\mathrm{tag}}| > 0.6$,
$\mathcal{D}$ is fixed to 0.6.

The OST was applied to the $B^0_s \rightarrow \mu^+ D_s^- X$
data sample,
yielding 
$N_{\mathrm{tag}} = 5601 \pm 102 \thinspace \mathrm{(stat)}$ 
candidates having an identified
initial state flavor, as shown in Fig.~\ref{prl_fig1}(b).  The tagging
efficiency was $(20.9\pm0.7)$\%.

After flavor tagging, the proper decay time of candidates is needed;
however,
the undetected neutrino and other missing particles in the
semileptonic $B^0_s$ decay prevent
a precise determination of the meson's momentum and Lorentz
boost. This represents an important contribution to the smearing of 
the proper decay length in semileptonic decays, in addition to the 
resolution effects. 
A correction factor $K$ was
estimated from a Monte Carlo (MC) simulation by finding
the distribution
of
$K = p_T(\mu^+ D_s^-)/p_T(B)$ for a given decay channel in bins of $M(\mu^+ D_s^-)$.
The proper decay length of each $B^0_s$ meson is then
$c t(B^0_s) = l_M K$, where 
$l_M = M(B^0_s) \cdot (\vec{L}_{T}\cdot
\vec{p}_T(\mu^+ \ds))/(p_{T}(\mu^+ \ds))^{2}$ is the measured
visible proper decay length (VPDL),
$\vec{L}_T$ is the vector from the PV 
to the $B^0_s$ decay vertex in the transverse plane and
$M(B^0_s) = 5.3696$~GeV$/c^2$~\cite{pdg}.

All flavor-tagged events with
$1.72 < M(K^+K^-\pi^-) < 2.22$~GeV$/c^2$ were used in an
unbinned fitting procedure. The likelihood, $\mathcal{L}$, 
for an event to
arise from a specific source in the sample depends 
event-by-event 
on $l_M$,
its uncertainty $\sigma_{l_M}$, the invariant mass
of the candidate $M(K^+K^- \pi^-)$, the predicted
dilution $\mathcal{D}(d_{\mathrm{tag}})$, and
the  selection variable $y_{\mathrm{sel}}$.
The {\sl pdf}s for $\sigma_{l_M}$, $M(K^+K^- \pi^-)$, $\mathcal{D}(d_{\mathrm{tag}})$
and $y_{\mathrm{sel}}$
were determined from data.
Four sources were considered:
the signal $\mu^+ D_s^-(\rightarrow \phi \pi^-)$;
the accompanying peak due to $\mu^+ D^{-} (\rightarrow \phi \pi^-)$;
a small (less than 1\%) reflection  
due to $\mu^+ D^{-} (\rightarrow K^{+} \pi^{-} \pi^-)$, where
the kaon mass is misassigned to one of the pions;
and combinatorial background.
The total fractions of the first two categories were 
determined from the mass fit of Fig.~\ref{prl_fig1}(b).

The $\mu^+ D_s^-$ signal sample is composed mostly 
of $B^0_s$ mesons with some contributions 
from $B^0_d$ and
$B^+$ mesons. Contributions of 
$b$ baryons to the sample were estimated to be small
and were neglected. The data were divided into
subsamples with and without oscillation as determined by
the OST. The distribution of the VPDL
 $l$ for non-oscillated and oscillated cases
was modeled appropriately for each type of $B$ meson, e.g., for
$B^0_s$:
\begin{eqnarray}
\label{pnososc}
\lefteqn{p_s^{\mathrm{nos/osc}}(l,K,d_{\mathrm{tag}}) = } \\ \nonumber
& & 
\frac{K}{c \tau_{B_s^0}} \exp(- \frac{Kl}{c \tau_{B_s^0}})
[1 \pm \mathcal{D}(d_{\mathrm{tag}})
\cos(\Delta m_s \cdot Kl/c)]/2.
\end{eqnarray}
The world averages~\cite{pdg} of $\tau_{B^0_d}$, $\tau_{B^+}$, and
$\Delta m_d$ were used as inputs to the fit.  The lifetime,
$\tau_{B^0_s}$, was allowed to float in the fit. In the amplitude and
likelihood scans described below, $\tau_{B^0_s}$ was fixed to this fitted
value, which agrees with 
expectations.

The total VPDL
{\sl pdf} for the $\mu^+ D_s^-$ signal is then the sum over all 
decay channels, including branching fractions,
that yield the $D_s^-$ mass peak.
The $B^0_s \to \mu^+ D_s^- X$ signal modes (including 
$D^{*-}_s$, ${D}_{s0}^{*-}$, and ${D}_{s1}^{'-}$; and
$\mu^+$ originating from
$\tau^+$ decay)
comprise $(85.6 \pm 3.3)$\% of our sample,
as determined from a MC
simulation which included
the {\sc PYTHIA} generator v6.2~\cite{pythia} interfaced with the
{\sc EVTGEN} decay package~\cite{evtgen}, followed by full
{\sc GEANT} v3.15~\cite{geant}
 modeling of the detector response and event reconstruction.
Other backgrounds considered were
decays via 
$B^0_s \to D^+_{(s)} D^-_s X$ and
$\bar{B}^0_d, B^- \to D D^-_s$, followed
by $ D^+_{(s)} \to \mu^+ X$, with 
a real $D_s^-$ reconstructed in the peak and an
associated real $\mu^+$.
Another background taken into account occurs
when the $D_s^-$ meson originates from one $b$ or $c$ quark and the
muon arises from another quark. This background peaks around the 
PV (peaking backgrounds).
The uncertainty in each
channel covers possible trigger efficiency biases.
Translation from the true VPDL, $l$, to the 
measured $l_M$ for a given channel, is achieved
by a convolution of the  VPDL detector resolution,
of $K$ factors over each normalized distribution, 
and by including the reconstruction efficiency as a function of VPDL.
The lifetime-dependent efficiency was
found for each channel using MC simulations and, as a cross
check, the efficiency was also determined from the data
by fixing $\tau_{B^0_s}$ and fitting for the
functional form of the efficiency.
The shape of the VPDL distribution for peaking backgrounds
was found from
MC simulation, and the fraction from this source was allowed to
float in the fit.


The VPDL uncertainty was determined from the vertex fit using track
parameters and their uncertainties. To account for possible mismodeling
of these uncertainties, resolution scale factors were introduced as
determined by examining the pull distribution of the vertex positions of
a sample of $J/\psi \rightarrow \mu^+ \mu^-$ decays. Using these scale
factors, the convolving function for the VPDL resolution was the sum of
two Gaussians with widths (fractions) of 0.998$\sigma_{l_M}$ (72\%) and
1.775$\sigma_{l_M}$ (28\%). A cross check was performed using a MC
simulation with tracking errors tuned according to the procedure
described in~\cite{tune}. The 7\% variation of scale factors found in this
cross check was used to estimate systematic uncertainties due to decay
length resolution.

Several contributions to the combinatorial backgrounds that have 
different VPDL distributions
were considered.  True
prompt background was modeled with a Gaussian function
with a separate scale factor on the width;
background due to fake vertices around the PV 
was modeled with another Gaussian function;
and long-lived background was modeled with an exponential
function convoluted with the resolution, including
a component oscillating with a frequency 
of $\Delta m_d$. 
The unbinned fit of the total tagged sample was used 
to determine the various
fractions of signal and backgrounds and the background
VPDL parametrizations.  

\begin{figure}
\includegraphics[width=0.42\textwidth]{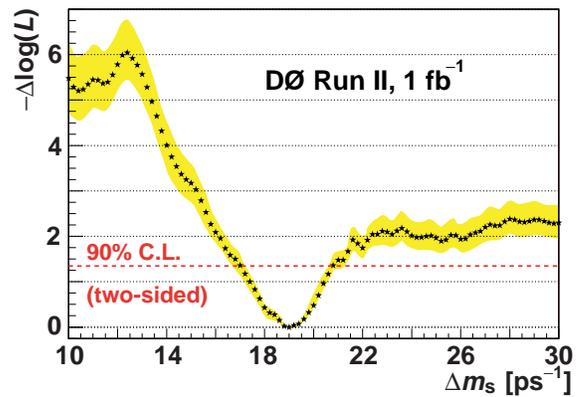}
\caption{\label{prl_fig2}
Value of $-\Delta\log\mathcal{L}$ as a function of $\Delta m_s$. 
Star symbols do not include systematic uncertainties, and the shaded 
band represents the envelope of all $\log\mathcal{L}$ scan curves
due to different systematic uncertainties.
}
\end{figure}

Figure~\ref{prl_fig2} shows the value of 
$-\Delta\log\mathcal{L}$ as a function of $\Delta m_s$, indicating
a favored value of 19~ps$^{-1}$, while variation of
$-\log\mathcal{L}$ from the minimum indicates an oscillation
frequency of $17 < \Delta m_s < 21$~ps$^{-1}$ at the 90\% C.L.
The uncertainties are approximately Gaussian inside this interval.
The plateau of the likelihood curve shows the region
where we do not have sufficient resolution to measure an oscillation,
and if the true value of $\Delta m_s > 22$~ps$^{-1}$, our measured
confidence interval does not make any statement about the frequency.
Using $100$ 
parametrized MC samples 
with similar statistics,
VPDL resolution, overall tagging performance, and
sample composition of the data
sample, it was determined that for a true value
of $\Delta m_s = 19$~ps$^{-1}$, the probability 
was 15\%
for measuring a value in the range $16 < \Delta m_s < 22$~ps$^{-1}$  
with a $-\Delta \log\mathcal{L}$ lower by at least 1.9 than the
corresponding value at $\Delta m_s = 25$~ps$^{-1}$.

\begin{figure}
\includegraphics[width=0.50\textwidth]{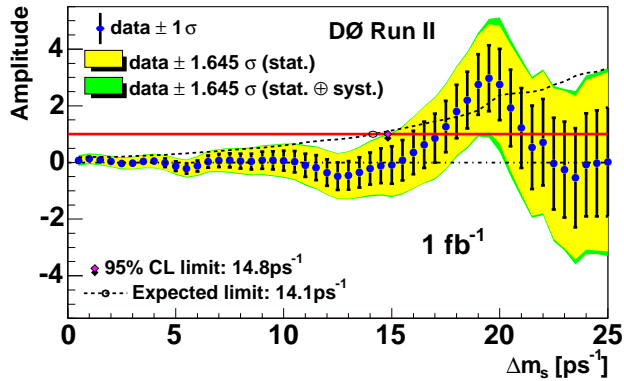}
\caption{\label{prl_fig3}
$B^0_s$ oscillation amplitude as a function of oscillation frequency,
$\Delta m_s$.  The solid line shows the $\mathcal{A}=1$ axis for
reference.  The dashed line shows the expected limit including both
statistical and systematic uncertainties.  }
\end{figure}

The amplitude method~\cite{amp_method} was also used.
Equation~\ref{pnososc} was modified to include
the oscillation amplitude $\mathcal{A}$ as an additional coefficient
on the $\cos(\Delta m_s\cdot Kl/c)$ term. The unbinned fit
was repeated for fixed input values of $\Delta m_s$ and
the fitted value of $\mathcal{A}$ and its uncertainty 
$\sigma_{\mathcal{A}}$ found for each step, as shown in 
Fig.~\ref{prl_fig3}.
At $\Delta m_s = 19$~ps$^{-1}$ the measured data point deviates from
the hypothesis $\mathcal{A}=0$ ($\mathcal{A}=1$) by 2.5 (1.6) standard
deviations, corresponding to a two-sided C.L. of 1\% (10\%), and is in
agreement with the likelihood results. 
In the presence of a signal, however, it is more difficult to define
a confidence interval using the amplitude than by examining the 
$-\Delta\log\mathcal{L}$
curve. Since, on average, these two methods give
the same results, we chose to quantify our $\Delta m_s$ 
interval using the likelihood curve.

Systematic uncertainties were addressed by varying inputs,
cut requirements,
branching ratios, and {\sl pdf} modeling. 
The branching ratios were varied within known uncertainties~\cite{pdg}
and large variations were taken for those not yet measured.
The $K$-factor distributions were varied within uncertainties,
using measured (or smoothed) instead of generated momenta 
in the MC simulation. The fractions of peaking and combinatorial 
backgrounds
were varied within uncertainties.  Uncertainties in the reflection
contribution were considered. The functional form to
determine the dilution $\mathcal{D}(d_{\mathrm{tag}})$ was varied.
The lifetime $\tau_{B^0_s}$ was fixed to its world average
value, and $\Delta \Gamma_s$
was allowed to be non-zero.
The scale factors on the signal and background resolutions 
were varied within uncertainties, and typically generated
the largest systematic uncertainty in the region of interest.
A separate scan of $-\Delta\log\mathcal{L}$ was taken for
each variation, and the envelope of all such curves is indicated
as the band in Fig.~\ref{prl_fig2}.  
The same systematic uncertainties were considered for
the amplitude method using the procedure of Ref.~\cite{amp_method}, 
and, when added in quadrature with the statistical
uncertainties, represent a small effect, as shown in Fig.~\ref{prl_fig3}.
Taking these systematic uncertainties into account,  
we obtain from
the amplitude method an expected limit of $14.1$~ps$^{-1}$ and
an observed lower limit of $\Delta m_s > 14.8$~ps$^{-1}$ at the 95\% C.L.,
consistent with the likelihood scan. 

The probability that $B^0_s$-$\bar{B}^0_s$ oscillations with the 
true value of $\Delta m_s > 22$~ps$^{-1}$
 would give a $-\Delta \log\mathcal{L}$ minimum in the
range $16 < \Delta m_s < 22$~ps$^{-1}$ with a depth
of more than 1.7 with respect to the $-\Delta \log\mathcal{L}$ 
value at $\Delta m_s = 25$~ps$^{-1}$, 
corresponding to our observation
including systematic uncertainties, was found to be $(5.0 \pm 0.3)\%$.
This range of $\Delta m_s$ was chosen to encompass the world average
lower limit and the edge of our sensitive region.
To determine this probability, an
ensemble test using the data sample was performed by randomly
assigning a flavor to each candidate while retaining all its other
information, effectively simulating a $B^0_s$
oscillation with an infinite frequency.  
Similar probabilities were found using ensembles of parametrized MC
events.
 

In summary, a study of $B^0_s$-$\bar{B}^0_s$ oscillations was
performed using $B^0_s \to \mu^+ D_s^- X$ decays, where $D_s^- \to
\phi \pi^-$ and $\phi \to K^+K^-$, an opposite-side flavor tagging
algorithm, and an unbinned likelihood fit. 
The amplitude method gives an expected limit of $14.1$~ps$^{-1}$ and
an observed lower limit of $\Delta m_s > 14.8$~ps$^{-1}$ at the 95\% C.L.
At $\Delta m_s =
19$~ps$^{-1}$, the amplitude method yields a result that
deviates from
the hypothesis $\mathcal{A}=0$ ($\mathcal{A}=1$) by 2.5 (1.6) 
standard deviations, corresponding to a two-sided C.L. of 1\% (10\%).
The likelihood curve is
well behaved near a preferred value of 19~ps$^{-1}$ with a 90\%
C.L. interval of $17 < \Delta m_s < 21$~ps$^{-1}$, assuming Gaussian
uncertainties.  The lower edge of the confidence level interval is near the
world average 95\% C.L. lower limit $\Delta m_s > 16.6$~ps$^{-1}$~\cite{hfag}.
Ensemble tests indicate that if $\Delta m_s$ lies above the sensitive
region, i.e., above approximately 22~ps$^{-1}$, there is a
$(5.0\pm0.3)$\% probability that it would produce a likelihood minimum
similar to the one observed in the interval $16 < \Delta m_s <
22$~ps$^{-1}$.  This is the first report of a direct two-sided bound 
measured by a single experiment on the 
$B^0_s$ oscillation frequency.



\input acknowledgement_paragraph_r2.tex   

\end{document}

%% file: list_of_authors_r2.tex
%
\author{                                                                      
V.M.~Abazov,$^{36}$                                                           
B.~Abbott,$^{76}$                                                             
M.~Abolins,$^{66}$                                                            
B.S.~Acharya,$^{29}$                                                          
M.~Adams,$^{52}$                                                              
T.~Adams,$^{50}$                                                              
M.~Agelou,$^{18}$                                                             
J.-L.~Agram,$^{19}$                                                           
S.H.~Ahn,$^{31}$                                                              
M.~Ahsan,$^{60}$
% ...
\\                                                                            
\vskip 0.30cm                                                                 
\centerline{(D\O\ Collaboration)}                                             
\vskip 0.30cm                                                                 
}                                                                             
\affiliation{                                                                 
\centerline{$^{1}$Universidad de Buenos Aires, Buenos Aires, Argentina}       
\centerline{$^{2}$LAFEX, Centro Brasileiro de Pesquisas F{\'\i}sicas,         
                  Rio de Janeiro, Brazil}                                     
\centerline{$^{3}$Universidade do Estado do Rio de Janeiro,                   
                  Rio de Janeiro, Brazil}                                     
\centerline{$^{4}$Instituto de F\'{\i}sica Te\'orica, Universidade            
                  Estadual Paulista, S\~ao Paulo, Brazil}                     
\centerline{$^{5}$University of Alberta, Edmonton, Alberta, Canada,           
                  Simon Fraser University, Burnaby, British Columbia, Canada,}
\centerline{York University, Toronto, Ontario, Canada, and                    
                  McGill University, Montreal, Quebec, Canada}                
\centerline{$^{6}$Institute of High Energy Physics, Beijing,                  
                  People's Republic of China}                                 
\centerline{$^{7}$University of Science and Technology of China, Hefei,       
                  People's Republic of China}                                 
\centerline{$^{8}$Universidad de los Andes, Bogot\'{a}, Colombia}             
\centerline{$^{9}$Center for Particle Physics, Charles University,            
                  Prague, Czech Republic}                                     
\centerline{$^{10}$Czech Technical University, Prague, Czech Republic}        
\centerline{$^{11}$Center for Particle Physics, Institute of Physics,         
                   Academy of Sciences of the Czech Republic,                 
                   Prague, Czech Republic}                                    
\centerline{$^{12}$Universidad San Francisco de Quito, Quito, Ecuador}        
\centerline{$^{13}$Laboratoire de Physique Corpusculaire, IN2P3-CNRS,         
                   Universit\'e Blaise Pascal, Clermont-Ferrand, France}      
\centerline{$^{14}$Laboratoire de Physique Subatomique et de Cosmologie,      
                   IN2P3-CNRS, Universite de Grenoble 1, Grenoble, France}    
\centerline{$^{15}$CPPM, IN2P3-CNRS, Universit\'e de la M\'editerran\'ee,     
                   Marseille, France}                                         
\centerline{$^{16}$IN2P3-CNRS, Laboratoire de l'Acc\'el\'erateur              
                   Lin\'eaire, Orsay, France}                                 
\centerline{$^{17}$LPNHE, IN2P3-CNRS, Universit\'es Paris VI and VII,         
                   Paris, France}                                             
\centerline{$^{18}$DAPNIA/Service de Physique des Particules, CEA, Saclay,    
                   France}                                                    
\centerline{$^{19}$IReS, IN2P3-CNRS, Universit\'e Louis Pasteur, Strasbourg,  
                    France, and Universit\'e de Haute Alsace,                 
                    Mulhouse, France}                                         
\centerline{$^{20}$Institut de Physique Nucl\'eaire de Lyon, IN2P3-CNRS,      
                   Universit\'e Claude Bernard, Villeurbanne, France}         
\centerline{$^{21}$III. Physikalisches Institut A, RWTH Aachen,               
                   Aachen, Germany}                                           
\centerline{$^{22}$Physikalisches Institut, Universit{\"a}t Bonn,             
                   Bonn, Germany}                                             
\centerline{$^{23}$Physikalisches Institut, Universit{\"a}t Freiburg,         
                   Freiburg, Germany}                                         
\centerline{$^{24}$Institut f{\"u}r Physik, Universit{\"a}t Mainz,            
                   Mainz, Germany}                                            
\centerline{$^{25}$Ludwig-Maximilians-Universit{\"a}t M{\"u}nchen,            
                   M{\"u}nchen, Germany}                                      
\centerline{$^{26}$Fachbereich Physik, University of Wuppertal,               
                   Wuppertal, Germany}                                        
\centerline{$^{27}$Panjab University, Chandigarh, India}                      
\centerline{$^{28}$Delhi University, Delhi, India}                            
\centerline{$^{29}$Tata Institute of Fundamental Research, Mumbai, India}     
\centerline{$^{30}$University College Dublin, Dublin, Ireland}                
\centerline{$^{31}$Korea Detector Laboratory, Korea University,               
                   Seoul, Korea}                                              
\centerline{$^{32}$SungKyunKwan University, Suwon, Korea}                     
\centerline{$^{33}$CINVESTAV, Mexico City, Mexico}                            
\centerline{$^{34}$FOM-Institute NIKHEF and University of                     
                   Amsterdam/NIKHEF, Amsterdam, The Netherlands}              
\centerline{$^{35}$Radboud University Nijmegen/NIKHEF, Nijmegen, The          
                  Netherlands}                                                
\centerline{$^{36}$Joint Institute for Nuclear Research, Dubna, Russia}       
\centerline{$^{37}$Institute for Theoretical and Experimental Physics,        
                   Moscow, Russia}                                            
\centerline{$^{38}$Moscow State University, Moscow, Russia}                   
\centerline{$^{39}$Institute for High Energy Physics, Protvino, Russia}       
\centerline{$^{40}$Petersburg Nuclear Physics Institute,                      
                   St. Petersburg, Russia}                                    
\centerline{$^{41}$Lund University, Lund, Sweden, Royal Institute of          
                   Technology and Stockholm University, Stockholm,            
                   Sweden, and}                                               
\centerline{Uppsala University, Uppsala, Sweden}                              
\centerline{$^{42}$Physik Institut der Universit{\"a}t Z{\"u}rich,            
                   Z{\"u}rich, Switzerland}                                   
\centerline{$^{43}$Lancaster University, Lancaster, United Kingdom}           
\centerline{$^{44}$Imperial College, London, United Kingdom}                  
\centerline{$^{45}$University of Manchester, Manchester, United Kingdom}      
\centerline{$^{46}$University of Arizona, Tucson, Arizona 85721, USA}         
\centerline{$^{47}$Lawrence Berkeley National Laboratory and University of    
                   California, Berkeley, California 94720, USA}               
\centerline{$^{48}$California State University, Fresno, California 93740, USA}
\centerline{$^{49}$University of California, Riverside, California 92521, USA}
\centerline{$^{50}$Florida State University, Tallahassee, Florida 32306, USA} 
\centerline{$^{51}$Fermi National Accelerator Laboratory,                     
            Batavia, Illinois 60510, USA}                                     
\centerline{$^{52}$University of Illinois at Chicago,                         
            Chicago, Illinois 60607, USA}                                     
\centerline{$^{53}$Northern Illinois University, DeKalb, Illinois 60115, USA} 
\centerline{$^{54}$Northwestern University, Evanston, Illinois 60208, USA}    
\centerline{$^{55}$Indiana University, Bloomington, Indiana 47405, USA}       
\centerline{$^{56}$University of Notre Dame, Notre Dame, Indiana 46556, USA}  
\centerline{$^{57}$Purdue University Calumet, Hammond, Indiana 46323, USA}    
\centerline{$^{58}$Iowa State University, Ames, Iowa 50011, USA}              
\centerline{$^{59}$University of Kansas, Lawrence, Kansas 66045, USA}         
\centerline{$^{60}$Kansas State University, Manhattan, Kansas 66506, USA}     
\centerline{$^{61}$Louisiana Tech University, Ruston, Louisiana 71272, USA}   
\centerline{$^{62}$University of Maryland, College Park, Maryland 20742, USA} 
\centerline{$^{63}$Boston University, Boston, Massachusetts 02215, USA}       
\centerline{$^{64}$Northeastern University, Boston, Massachusetts 02115, USA} 
\centerline{$^{65}$University of Michigan, Ann Arbor, Michigan 48109, USA}    
\centerline{$^{66}$Michigan State University,                                 
            East Lansing, Michigan 48824, USA}                                
\centerline{$^{67}$University of Mississippi,                                 
            University, Mississippi 38677, USA}                               
\centerline{$^{68}$University of Nebraska, Lincoln, Nebraska 68588, USA}      
\centerline{$^{69}$Princeton University, Princeton, New Jersey 08544, USA}    
\centerline{$^{70}$State University of New York, Buffalo, New York 14260, USA}
\centerline{$^{71}$Columbia University, New York, New York 10027, USA}        
\centerline{$^{72}$University of Rochester, Rochester, New York 14627, USA}   
\centerline{$^{73}$State University of New York,                              
            Stony Brook, New York 11794, USA}                                 
\centerline{$^{74}$Brookhaven National Laboratory, Upton, New York 11973, USA}
\centerline{$^{75}$Langston University, Langston, Oklahoma 73050, USA}        
\centerline{$^{76}$University of Oklahoma, Norman, Oklahoma 73019, USA}       
\centerline{$^{77}$Oklahoma State University, Stillwater, Oklahoma 74078, USA}
\centerline{$^{78}$Brown University, Providence, Rhode Island 02912, USA}     
\centerline{$^{79}$University of Texas, Arlington, Texas 76019, USA}          
\centerline{$^{80}$Southern Methodist University, Dallas, Texas 75275, USA}   
\centerline{$^{81}$Rice University, Houston, Texas 77005, USA}                
\centerline{$^{82}$University of Virginia, Charlottesville,                   
            Virginia 22901, USA}                                              
\centerline{$^{83}$University of Washington, Seattle, Washington 98195, USA}  
}                                                                             

%% file: acknowledgement_paragraph_r2.tex
%
We thank the staffs at Fermilab and collaborating institutions, 
and acknowledge support from the 
DOE and NSF (USA);
CEA and CNRS/IN2P3 (France);
FASI, Rosatom and RFBR (Russia);
CAPES, CNPq, FAPERJ, FAPESP and FUNDUNESP (Brazil);
DAE and DST (India);
Colciencias (Colombia);
CONACyT (Mexico);
KRF and KOSEF (Korea);
CONICET and UBACyT (Argentina);
FOM (The Netherlands);
PPARC (United Kingdom);
MSMT (Czech Republic);
CRC Program, CFI, NSERC and WestGrid Project (Canada);
BMBF and DFG (Germany);
SFI (Ireland);
The Swedish Research Council (Sweden);
Research Corporation;
Alexander von Humboldt Foundation;
and the Marie Curie Program.

%% file: prl_bsm_v5_0.bbl
\begin{thebibliography}{99}
\bibitem{charge_conj}
Charge conjugate states are assumed throughout.
\end{thebibliography}
